\documentclass[a4paper,11pt]{article}
\usepackage[utf8]{inputenc}      
\usepackage[T1]{fontenc}
\usepackage{lmodern}
\usepackage{geometry}
\usepackage{amsmath,amssymb}
\usepackage{graphicx}
\usepackage{booktabs}
\usepackage{siunitx}             
\usepackage{hyperref}
\usepackage{authblk}
\allowdisplaybreaks
\usepackage{pdflscape}
\usepackage{rotating}
\usepackage{verbatim}

\title{Branching Fractions and CP-Violating Asymmetries in $D \to PP$ Decays}

\author[1]{Franco Buccella\thanks{buccella@ba.infn.it}}
\author[2,3]{Pietro Santorelli\thanks{pietro.santorelli@unina.it}}

\affil[1]{INFN Sezione di Bari, Bari, Italy}
\affil[2]{Dipartimento di Fisica “E. Pancini”, Universit\`{a} di Napoli Federico II, via Cintia, 80126 Napoli, Italy}
\affil[3]{INFN Sezione di Napoli, Napoli, Italy}

\begin{document}
\maketitle

\begin{abstract}
We update an \(SU(3)_F\) analysis of \(D\to PP\) decays ($P\equiv \pi, K$) in which the dominant 
\(SU(3)_F\)-breaking 
effects are modeled through final-state interactions.
The new fit improves the description relative to the previous parameter set, although significant 
global tension remains, 
mainly in charged-$D$ modes. Within the adopted rescattering model, the direct CP asymmetries 
in $D^0\to\pi^+\pi^-$, $K^+K^-$, $\pi^0\pi^0$, 
and $K_S^0 K_S^0$ depend on two additional reduced amplitudes. Using the $D^0\to\pi^+\pi^-$ and
$D^0\to K^+K^-$ asymmetries as inputs, 
we derive conditional predictions for the two neutral modes.
The OZI-motivated benchmark $\Delta_4=0$ is mildly disfavored by the charged-mode asymmetries and
leads to substantially
smaller asymmetries in the neutral modes, which can be tested with more precise $K_S^0K_S^0$ and 
$\pi^0\pi^0$ measurements.
\end{abstract}


\section{Introduction}
\label{sec:intro}

After a challenging experimental effort that provided observation of CP 
violation in charm decays at more than five standard deviations~\cite{LHCb:2019}
\begin{equation}
\Delta A_{CP} = -(1.54 \pm 0.29)\times 10^{-3} \,,
\end{equation}
the same collaboration has recently reported new measurements of
CP–violating asymmetries in the decays of the neutral charmed pseudoscalar 
meson, $D^{0}$, into two charged pions, two charged kaons, and two neutral kaons.
The extracted direct CP asymmetries in the charged final states
and the time-integrated asymmetry in the $K_S^0K_S^0$ final state
are~\cite{LHCb:2022,LHCb:2021rdn}
\begin{align}
A_{CP}(D^0 \to \pi^+ \pi^-) &= (+2.32 \pm 0.61)\times 10^{-3},
\label{e:LHCbpipi}\\
A_{CP}(D^0 \to K^+ K^-)     &= (+0.77 \pm 0.57)\times 10^{-3},
\label{e:LHCbKK}\\
A_{CP}(D^0 \to K_S^0 K_S^0)     &= (-31 \pm 12 \pm 4 \pm 2)\times 10^{-3}.
\end{align}
More recently, the LHCb Collaboration has reported an updated measurement of 
the time-integrated CP asymmetry in the $D^{0} \to K^{0}_{S} K^{0}_{S}$ decay
mode~\cite{LHCb:2025}
\begin{equation}
A_{CP}(D^{0} \to K^{0}_{S} K^{0}_{S}) = (18.6 \pm 10.4 \pm 4.1)\times 10^{-3}
\,.
\end{equation}
Combining this result with the previous LHCb measurements~\cite{LHCb:2015ope,LHCb:2021rdn}, 
and assuming that all systematic uncertainties are uncorrelated, one obtains
\begin{align}
A_{CP}(D^{0} \to K^{0}_{S} K^{0}_{S}) &= (-3.7 \pm 7.8 \pm 2.9)
\times 10^{-3}\,.
\end{align}
\noindent
The most recent measurements performed by Belle and Belle~II Collaborations 
are~\cite{Belle:2024vho,Belle:2025cud}:
\begin{align}
A_{CP}(D^{0} \to K^{0}_{S} K^{0}_{S}) &= (-14 \pm 13 \pm 1)\times 10^{-3},\\
A_{CP}(D^{0} \to K^{0}_{S} K^{0}_{S}) &= (+13 \pm 20 \pm 2)\times 10^{-3}.
\end{align}
which, when combined, yield
\begin{align}
A_{CP}(D^{0} \to K^{0}_{S} K^{0}_{S}) &= (-6.0 \pm 10.9 \pm 0.9)\times 10^{-3}.
\end{align}
The CMS Collaboration has also reported an independent measurement of this 
asymmetry~\cite{CMS:2024}:
\begin{equation}
A_{CP}(D^{0}\to K^{0}_{S}K^{0}_{S}) = (+62\pm30\pm2\pm8)\times10^{-3}\,.
\end{equation}
The first two uncertainties are statistical and systematic, respectively, while the third arises 
from the uncertainty on the CP asymmetry of the normalization channel \(D^{0}\to K^{0}_{S}\pi^{+}\pi^{-}\).
The CMS, LHCb, and Belle/Belle~II determinations are mutually compatible within their present uncertainties 
and show no significant evidence for CP violation in this decay mode. By averaging the most recent 
CMS, LHCb and Belle~II determinations we obtain:
\begin{equation}
A_{CP}(D^0 \to K^{0}_{S}K^{0}_{S})\big|_{\mathrm{ave}} = (-1.7 \pm 6.2 \pm 1.8) 
\times 10^{-3}.
\end{equation}
A complementary measurement in a neutral final state has recently been
reported by the Belle~II Collaboration~\cite{Belle-II:pi0pi0}:
\begin{equation}
A_{CP}(D^0\to\pi^0\pi^0)= (3.0\pm7.2\pm2.0)\times10^{-3}\,,
\label{eq:BelleII-pi0pi0}
\end{equation}
where, as usual, the first uncertainty is statistical and the second systematic.

\noindent
CP-violating asymmetries were advocated many years ago~\cite{BLMPS} 
as a consequence of the final-state interactions needed to account for the large $SU(3)_F$ 
violations in the amplitudes for charmed-particle decays into two pseudoscalar mesons 
and implied by the isospin identities for the Cabibbo-allowed amplitudes of $D$ decays into 
$\bar K\pi$~\cite{LOR} and for the singly Cabibbo-suppressed decays into two pions~\cite{FMS}.

Related resonance-dominated and coupled-channel descriptions of
$D\to PP$ decays were developed in
Refs.~\cite{BuccellaEtAl1996,BiswasEtAl2015}.
More recently, the role of final-state interactions in charm CP violation
has been studied through nearby scalar resonances~\cite{SchachtSoni2022},
CPT-constrained $\pi\pi\leftrightarrow K\bar K$
rescattering~\cite{BediagaFredericoMagalhaes2023}, and coupled-channel
dispersive analyses~\cite{PichSolomonidiValeSilva2023,PichSolomonidiValeSilva2026}.
For a recent review of charm physics, including its theoretical
and experimental status and CP violation, see
Ref.~\cite{Friday:2025gpj}.

In a previous paper~\cite{BPS}, we wrote the 
amplitudes for the decays of the charmed pseudoscalar mesons into two 
pseudoscalars, assuming $SU(3)_F$ for the non-rescattered 
amplitudes and a final-state interaction coming from the 
presence of a scalar nonet in the $D$-meson mass region 
necessary to explain the large phase differences between 
the isospin final states shown by the triangular isospin identities 
for the Cabibbo-favoured decays into $\bar{K} \pi$ \cite{LOR} and 
singly Cabibbo-suppressed decays into $\pi \pi$ \cite{FMS}. 
We also included $SU(3)_F$ breaking associated with the nonconservation of
strangeness–changing vector currents and introduced two additional
$SU(3)_F$-violating parameters, since the experimental precision on
branching fractions exceeds the nominal accuracy of exact $SU(3)_F$.

The leading reduced-amplitude structure remains stable after refitting, 
although the previous parameter set no longer provides an adequate description of 
several updated branching-fraction measurements.

To evaluate CP-violating asymmetries, one must compute the $\Delta U=0$
amplitudes arising from penguin and pseudo-penguin topologies, both
proportional to $V^{*}_{cb}V_{ub}$. These connect $D^0$ to the $U$-spin
singlet components within the $\mathbf{1}$, $\mathbf{8}$, and $\mathbf{27}$
representations of $SU(3)_F$.

Since the reduced matrix element of the tensor product 
$\mathbf{\bar{3}}\otimes\mathbf{15} \supset \mathbf{27}$ 
is determined from the $\Delta U = 1$ amplitudes, only two independent 
parameters are required to describe the $\Delta U = 0$ transitions.
The phases and the singlet–octet mixing of the isospin–zero states,
which generate the final–state interactions,
are also constrained by the analysis of the $\Delta U = 1$ matrix elements.\\
In summary, the $\Delta U = 0$ amplitudes and the corresponding CP-violating 
asymmetries depend on only two parameters, which in turn allow us to deduce 
the asymmetries into two $K_S$ and into two $\pi^0$ from 
the ones into $\pi^+\, \pi^-$ and $ K^+ K^-$.
If the Zweig selection rule~\cite{Okubo:1963fa,Zweig:1964jf,Iizuka:1966fk} is assumed, these asymmetries would 
depend on a single parameter.
 In the following section, we compare the parametrization developed in our 
previous paper~\cite{BPS} with the most recent experimental data~\cite{PDG2026}. 
We determine the parameter values that yield an improved agreement with the 
data while preserving similar values of several leading amplitude parameters.
 
In Sec.~\ref{sec:DeltaU0}, we derive the $\Delta U = 0$ amplitudes for the neutral $D$ 
decays, assuming $SU(3)_F$ symmetry for the non-rescattered amplitudes 
and adopting the pattern of final-state interactions that reproduces the 
observed branching fractions. 
Within this framework, we express the CP-violating asymmetries in terms of two 
parameters, which reduce to a single one if the Zweig selection rule~\cite{Okubo:1963fa,Zweig:1964jf,Iizuka:1966fk} 
is assumed, as discussed in Sec.~\ref{sec:OZI-benchmark}. Finally, Sec.~\ref{sec:conclusions} summarizes our conclusions.

\section{The branching ratios into two pseudoscalar mesons}
\label{sec:BR}

Nonleptonic $D\to PP$ decays exhibit sizable $SU(3)_F$ breaking,
while the isospin triangle relations for the Cabibbo-allowed
$D\to\bar K\pi$~\cite{LOR} and singly Cabibbo-suppressed
$D\to\pi\pi$~\cite{FMS} modes require different strong phases
in the final isospin channels. Ref.~\cite{BLMPS} relates these
observations to a scalar-meson nonet near the $D$-meson scale,
with large variations in the rescattering phases associated with
Gell-Mann--Okubo mass splittings and isoscalar singlet--octet mixing.
The subsequent analyses of neutral-$D$ SCS decays~\cite{BLPS} and
of the complete $D\to PP$ system~\cite{BPS} assume that the dominant
$SU(3)_F$ breaking arises from FSI and the nonconservation of
strangeness-changing vector currents.

The Cabibbo-allowed, singly Cabibbo-suppressed, and doubly
Cabibbo-suppressed amplitudes are proportional to the matrix
elements of the $\Delta U=1$ operators
\begin{align}
&\bar{u}_L(x)\gamma_{\mu} d_L(x)\,\bar{s}_L(x)\gamma^{\mu} c_L(x), 
\label{equation7}\\
&\bar{u}_L(x)\gamma_{\mu} s_L(x)\,\bar{s}_L(x)\gamma^{\mu} c_L(x)
- \bar{u}_L(x)\gamma_{\mu} d_L(x)\,\bar{d}_L(x)\gamma^{\mu} c_L(x), 
\label{equation8}\\
&\bar{u}_L(x)\gamma_{\mu} s_L(x)\,\bar{d}_L(x)\gamma^{\mu} c_L(x), 
\label{equation9}
\end{align}
and to the CKM combinations $V^{*}_{cs}V_{ud}$,
$(V^{*}_{cs}V_{us}-V^{*}_{cd}V_{ud})/2$, and
$-V^{*}_{cd}V_{us}$, respectively. The four neutral-$D$
$\Delta U=1$ amplitudes relevant to the subsequent CP analysis
are~\cite{BPS}
\begin{align}
A(D^0\to\pi^+\pi^-) &= \left(T - \frac{2}{3} C\right) \left[-\frac{3}{10} 
\left(e^{i \delta_0} + e^{i \delta_0^\prime}\right) + \left(-\frac{3}{10}  
\cos(2 \phi) + \frac{3}{4\sqrt{10}} \sin(2 \phi)\right)  \left(e^{i 
\delta_0^\prime} - e^{i \delta_0}\right)\right]\nonumber\\& -  
\frac{2}{5}\left(T + C\right)\nonumber\\
A(D^0\to K^+K^-) &= \left(T - \frac{2}{3} C\right) \left[\frac{3}{20} 
\left(e^{i \delta_0} + e^{i \delta_0^\prime}\right) + \left(\frac{3}{20}  
\cos(2 \phi) + \frac{3}{4\sqrt{10}} \sin(2 \phi)\right)  
\left(e^{i \delta_0^\prime} - e^{i \delta_0}\right)\right.\nonumber\\&\left. 
+ \frac{3}{10} e^{i \delta_1}\right] +  \frac{2}{5}\left(T + C
\right)\nonumber\\
A(D^0\to K^0\bar{K}^0) &= \left(T - \frac{2}{3} C\right) \left[\frac{3}{20} 
\left(e^{i \delta_0} + e^{i \delta_0^\prime}\right) + \left(\frac{3}{20}  
\cos(2 \phi) + \frac{3}{4\sqrt{10}} \sin(2 \phi)\right)  
\left(e^{i \delta_0^\prime} - e^{i \delta_0}\right)
\right.\nonumber\\& \left.-  \frac{3}{10} e^{i \delta_1}\right]\nonumber\\
A(D^0\to\pi^0\pi^0) &= \left(T - \frac{2}{3} C\right) \left[-\frac{3}{10} \left(e^{i \delta_0} + e^{i \delta_0^\prime}\right) + 
\left(-\frac{3}{10}  \cos(2 \phi) + \frac{3}{4\sqrt{10}} \sin(2 \phi)\right)  \left(e^{i \delta_0^\prime} - e^{i \delta_0}\right)\right]
\nonumber\\& +  \frac{3}{5}\left(T + C\right)\,.
\end{align}
The remaining amplitudes are reported in
Appendix~\ref{sec:DeltaU=1A}.

The $\Delta S=0$ amplitudes also receive contributions from the
pseudopenguin and penguin operators
\begin{equation}
\bar{u}_L(x) \gamma_{\mu} s_L(x) \bar{s}_L(x) \gamma^{\mu} c_L(x) +
\bar{u}_L(x) \gamma_{\mu} d_L(x) \bar{d}_L(x) \gamma^{\mu} c_L(x)
\label{equation410}
\end{equation}
\begin{equation}
\left[\bar{u}(x) \lambda_a \gamma_{\mu} u(x) +\
\bar{d}(x) \lambda_a \gamma_{\mu} d(x) +
\bar{s}(x) \lambda_a \gamma_{\mu} s(x)
\right] \,\bar{u}_L(x) \lambda_a \gamma^{\mu} c_L(x) 
\label{equation511}
\end{equation}
both with $\Delta U=0$ and proportional to $-V^*_{cb}V_{ub}/2$.

\begin{table}[htbp]
\centering
\renewcommand{\arraystretch}{1.10}
\setlength{\tabcolsep}{6pt}
\begin{tabular}{l c c @{\hspace{0.8cm}} l c c}
\toprule
\cmidrule(lr){1-3}\cmidrule(l){4-6}
Par. & Old $(\mu \pm \sigma)$ & Present $(\mu \pm \sigma)$
& Par. & Old $(\mu \pm \sigma)$ & Present $(\mu \pm \sigma)$ \\
\midrule

$T$
& $\num{0.424} \pm \num{0.003}$
& $\num{0.4261} \pm \num{0.0020}$
&
$\delta_0$
& $\num{-2.373} \pm \num{0.062}$
& $\num{-2.633} \pm \num{0.067}$ \\

$C$
& $\num{-0.211} \pm \num{0.003}$
& $\num{-0.2108} \pm \num{0.0027}$
&
$\delta'_0$
& $\num{-0.840} \pm \num{0.046}$
& $\num{-1.021} \pm \num{0.029}$ \\

$\kappa$
& $\num{-0.036} \pm \num{0.004}$
& $\num{-0.0378} \pm \num{0.0034}$
&
$\delta_{1/2}$
& $\num{-1.632} \pm \num{0.020}$
& $\num{-1.632} \pm \num{0.017}$ \\

$\kappa^\prime$
& $\num{-0.063} \pm \num{0.088}$
& $\num{-0.076} \pm \num{0.074}$
&
$\delta_1$
& $\num{-1.085} \pm \num{0.038}$
& $\num{-1.221} \pm \num{0.031}$ \\

$K$
& $\num{0.100} \pm \num{0.012}$
& $\num{0.0969} \pm \num{0.0094}$
&
& & \\

$K^\prime$
& $\num{-0.153} \pm \num{0.072}$
& $\num{0.167} \pm \num{0.032}$
&
& & \\

$\Delta$
& $\num{-0.026} \pm \num{0.019}$
& $\num{0.035} \pm \num{0.017}$
&
& & \\

$\phi$
& $\num{0.435} \pm \num{0.025}$
& $\num{0.350} \pm \num{0.019}$
&
& & \\

$\epsilon_\delta$
& $\num{0.067} \pm \num{0.061}$
& $\num{0.168} \pm \num{0.031}$
&
& & \\

\bottomrule
\end{tabular}
\caption{
Parameter comparison between the old fit~\cite{BPS} and the
present fit. Present uncertainties are obtained by symmetrizing
the MINOS errors, $\Delta_\pm$, as
$\sigma_{\rm sym}
=\tfrac{1}{2}\left(|\Delta_-|+|\Delta_+|\right)$.
For the present fit, uncertainties are rounded to two significant
digits and central values to the corresponding decimal place.
The parameters $T$, $C$, $\Delta$, $\kappa^{(\prime)}$, and
$K^{(\prime)}$ are in units of $\mathrm{GeV}^3$.
The angle $\phi$ and the phases $\delta_0$, $\delta'_0$,
$\delta_{1/2}$, and $\delta_1$ are in radians.
The parameter $\epsilon_\delta$ is dimensionless.
}
\label{t:params_old_vs_present}
\end{table}
\begin{table}[h]
\centering
\scriptsize
\renewcommand{\arraystretch}{1.12}
\setlength{\tabcolsep}{4pt}

\begin{tabular}{lccc@{\hspace{0.30cm}}ccc@{\hspace{0.30cm}}c}
\hline\hline
\cline{2-4}\cline{5-7}\cline{8-8}
Channel
& Old fit & PDG (2016) & PDG (2026)
& Present fit & pull$_{\rm Present}$ & pull$_{\rm Old,\,2016}$
& pull$_{\rm Old,\,2026}$ \\
\hline
\multicolumn{8}{c}{\textbf{SCS}}\\

$D^{0}\to\pi^{+}\pi^{-}$
& $1.448 \pm 0.019$ & $1.407 \pm 0.025$ & $1.451 \pm 0.024$
& $1.445 \pm 0.025$ & $+0.25$ & $-1.64$
& $+0.13$ \\

$D^{0}\to\pi^{0}\pi^{0}$
& $0.816 \pm 0.025$ & $0.822 \pm 0.025$ & $0.825 \pm 0.025$
& $0.816 \pm 0.025$ & $+0.36$ & $+0.24$
& $+0.36$ \\

$D^{+}\to\pi^{+}\pi^{0}$
& $1.235 \pm 0.033$ & $1.17 \pm 0.06$ & $1.243 \pm 0.033$
& $1.253 \pm 0.033$ & $-0.30$ & $-1.08$
& $+0.24$ \\

$D^{0}\to K^{+}K^{-}$
& $4.064 \pm 0.044$ & $3.97 \pm 0.07$ & $4.070 \pm 0.060$
& $3.955 \pm 0.050$ & $+1.92$ & $-1.34$
& $+0.10$ \\

$D^{0}\to K^{0}_{S}K^{0}_{S}$
& $0.168 \pm 0.012$ & $0.17 \pm 0.012$ & $0.1410 \pm 0.0050$
& $0.1406 \pm 0.0050$ & $+0.08$ & $+0.17$
& $-5.40$ \\

$D^{+}\to K^{+}K^{0}_{S}$
& $3.164 \pm 0.056$ & $2.83 \pm 0.16$ & $3.020 \pm 0.090$
& $3.148 \pm 0.072$ & $-1.42$ & $-2.09$
& $-1.60$ \\

$D^{+}\to K^{+}K^{0}_{L}$
& $3.164 \pm 0.056$ & -- & $3.21 \pm 0.16$
& $3.148 \pm 0.072$ & $+0.39$ & --
& $+0.29$ \\

$D_{s}^{+}\to\pi^{0}K^{+}$
& $1.41 \pm 0.15$ & $0.63 \pm 0.21$ & $0.750 \pm 0.050$
& $0.830 \pm 0.047$ & $-1.60$ & $-3.71$
& $-13.20$ \\

$D_{s}^{+}\to\pi^{+}K^{0}_{S}$
& $1.24 \pm 0.06$ & $1.22 \pm 0.06$ & $1.220 \pm 0.040$
& $1.225 \pm 0.040$ & $-0.13$ & $-0.33$
& $-0.50$ \\

\hline
\multicolumn{8}{c}{\textbf{CA and DCS}}\\

$D^{+}\to\pi^{+}K^{0}_{S}$
& $15.80 \pm 0.29$ & $14.7 \pm 0.8$ & $15.54 \pm 0.33$
& $15.62 \pm 0.31$ & $-0.24$ & $-1.38$
& $-0.79$ \\

$D^{+}\to\pi^{+}K^{0}_{L}$
& $14.37 \pm 0.52$ & $14.6 \pm 0.5$ & $14.60 \pm 0.50$
& $14.32 \pm 0.45$ & $+0.56$ & $+0.46$
& $+0.46$ \\

$D^{0}\to\pi^{+}K^{-}$
& $38.96 \pm 0.32$ & $38.9 \pm 0.4$ & $39.36 \pm 0.30$
& $39.477 \pm 0.299$ & $-0.39$ & $-0.15$
& $+1.33$ \\

$D^{0}\to\pi^{0}K^{0}_{S}$
& $12.29 \pm 0.21$ & $11.9 \pm 0.4$ & $12.39 \pm 0.22$
& $12.43 \pm 0.19$ & $-0.18$ & $-0.98$
& $+0.45$ \\

$D^{0}\to\pi^{0}K^{0}_{L}$
& $9.73 \pm 0.21$ & $10.0 \pm 0.7$ & $9.76 \pm 0.32$
& $9.86 \pm 0.16$ & $-0.31$ & $+0.39$
& $+0.09$ \\

$D^{+}\to\pi^{0}K^{+}$
& $0.151 \pm 0.013$ & $0.181 \pm 0.027$ & $0.154 \pm 0.013$
& $0.1068 \pm 0.0089$ & $+3.63$ & $+1.11$
& $+0.23$ \\

$D^{0}\to\pi^{-}K^{+}$
& $0.141 \pm 0.003$ & $0.1385 \pm 0.0027$ & $0.1359 \pm 0.0025$
& $0.1402 \pm 0.0026$ & $-1.72$ & $-0.93$
& $-2.04$ \\

$D_{s}^{+}\to K^{+}K^{0}_{S}$
& $14.67 \pm 0.41$ & $15.0 \pm 0.5$ & $15.00 \pm 0.14$
& $14.93 \pm 0.14$ & $+0.50$ & $+0.66$
& $+2.36$ \\

$D_{s}^{+}\to K^{+}K^{0}_{L}$
& -- & -- & $14.90 \pm 0.60$
& $15.47 \pm 0.29$ & $-0.95$ & --
& -- \\

\hline\hline
\multicolumn{5}{r}{\textbf{$\sum$ pull$^2$}}
& \textbf{26.66}
& \textbf{29.77}
& \textbf{219.10} \\
\hline\hline
\end{tabular}
\caption{Comparison of the experimental branching-ratio averages with the
predictions obtained from the old and present fits. For the old fit, the
table reports the predictions obtained with the parameter values of
Ref.~\cite{BPS}, together with their comparison with both the PDG(2016)
averages used in the original analysis and the updated PDG(2026) data. For
the present fit, the branching-ratio predictions are reported as the mean
values of the propagated fit distributions, with uncertainties given by
their standard deviations, and are compared with the PDG(2026) averages.
The pulls are defined as
\(
{\rm pull}
=
({\rm EXP}-\mu_{\rm fit})/\sigma_{\rm exp},
\)
where only the experimental uncertainty is included in the denominator.
All branching ratios are given in units of \(10^{-3}\).
\label{t:brfit_old_present}}
\end{table}

The $\Delta U=1$ amplitudes are expressed in terms of three real
parameters, $T$, $C$, and $\Delta$, associated with the reduced
matrix elements of
$\mathbf{\bar 3}\otimes\mathbf{\bar 6}\to\mathbf{8}$,
$\mathbf{\bar 3}\otimes\mathbf{15}\to\mathbf{8}$, and
$\mathbf{\bar 3}\otimes\mathbf{15}\to\mathbf{27}$.
FSI are described by the four phases $\delta_0$, $\delta'_0$,
$\delta_{1/2}$, and $\delta_1$, and the octet--singlet mixing
angle $\phi$ of the two isoscalar states. A common fractional
shift of the phases entering the $D_s^+$ amplitudes is parametrized by
\begin{equation}
\delta'_{1}=\delta_{1}(1-\epsilon_\delta),
\qquad
\delta'_{1/2}=\delta_{1/2}(1-\epsilon_\delta).
\end{equation}
The parameters $K$ and $K'$ account for $SU(3)_F$ breaking from
the nonconservation of strangeness-changing vector currents.
In the factorization interpretation, $K$ is associated with an
electrically neutral flavor current, whereas $K'$ enters
annihilation-type amplitudes of charged $D^+$ and $D_s^+$ mesons,
which require an electrically charged current to annihilate the
initial meson. Finally, $\kappa$ and $\kappa'$ parametrize
additional $SU(3)_F$ breaking in the $\mathbf{27}$ reduced
amplitude. Further details are given in Ref.~\cite{BPS}.

Table~\ref{t:params_old_vs_present} compares the old and present
parameter sets. The values of $T$, $C$, $\kappa$, $K$, and $\delta_{1/2}$
remain close to the previous results, while $\kappa'$ remains
weakly constrained and compatible with its previous value. The central values of
$K'$ and $\Delta$ change from negative to positive; $\delta_0$,
$\delta'_0$, and $\delta_1$ become more negative, $\phi$
decreases, and $\epsilon_\delta$ increases. The update therefore
mainly readjusts the subleading $SU(3)_F$-breaking and rescattering
contributions.
The branching-fraction fit has an exact twofold ambiguity under
the simultaneous reversal of all rescattering phases.
The positive- and negative-phase solutions therefore yield identical
branching-fraction predictions and values of $\chi^2$.
We adopt the negative-phase branch, which, within our phase
convention, is consistent with a scalar-resonance interpretation
of FSI. The CP predictions below are conditional on this choice.

Table~\ref{t:brfit_old_present} compares the old and present
branching-fraction predictions with the PDG(2016)~\cite{PDG:2016}
and PDG(2026)~\cite{PDG2026} averages. The 18 updated measurements
are fitted with 13 free hadronic parameters by minimizing
\begin{equation}
\chi^2_{\rm BR}\bigl(\theta_\alpha\bigr)=
\sum_{i=1}^{18}
\left[
\frac{
\mathrm{Br}^{\rm exp}_i-
\mathrm{Br}^{\rm th}_i\bigl(\theta_\alpha\bigr)
}{
\sigma^{\rm exp}_i
}
\right]^2,
\label{eq:chi2-BR}
\end{equation}
where $\theta_\alpha$ denotes the fitted parameters and
$\mathrm{Br}^{\rm exp}_i$ and $\sigma_i^{\rm exp}$ are the
PDG(2026) inputs in Table~\ref{t:brfit_old_present}.
The branching-fraction fit uses a diagonal experimental covariance
matrix; correlations between the experimental inputs are neglected.
The minimization uses \texttt{iminuit}, the Python interface to
MINUIT2~\cite{iminuit,MINUIT}, and the parameter uncertainties
are determined with MINOS.

The branching-fraction predictions in
Table~\ref{t:brfit_old_present} are the means and standard
deviations obtained by Monte Carlo propagation of the covariance
matrix at the minimum, including parameter correlations.
We define the pulls as
\begin{equation}
{\rm pull}_i=
\frac{
\mathrm{Br}^{\rm exp}_i-\mu_{{\rm fit},i}
}{
\sigma^{\rm exp}_i
},
\label{eq:pull-BR}
\end{equation}
using only the experimental uncertainties in the denominator.
Since $\mu_{{\rm fit},i}$ is a propagated mean rather than the
prediction evaluated at the minimum, $\sum_i{\rm pull}_i^2$
is a diagnostic quantity and need not equal $\chi^2_{\rm BR,min}$.

For comparison with the PDG(2026) inputs, the sum of squared pulls
changes from $219.10$ for the old parameter set to $26.66$ for
the present fit. The disagreement of the old predictions is
dominated by $D_s^+\to\pi^0K^+$ and $D^0\to K_S^0K_S^0$.
In the present fit, the largest residual occurs in
$D^+\to\pi^0K^+$, with ${\rm pull}\simeq+3.63$, whereas
all SCS branching fractions, including the neutral-$D$ channels
relevant to the CP analysis, are reproduced within $2\sigma$.

Despite this improvement, the full fit remains statistically
unsatisfactory: $\chi^2_{\rm BR,min}=26.80$ for five degrees
of freedom gives a nominal $p$-value of $6.2\times10^{-5}$.
Excluding $D^+\to\pi^0K^+$ and refitting the remaining
17 channels with the same 13 parameters yields
$\chi^2_{\rm BR,min}=8.17$ for four degrees of freedom,
with $p=0.086$. These $p$-values assume the usual chi-square
approximation and uncorrelated Gaussian experimental uncertainties.
The exclusion is a diagnostic stability check, not a reason to
discard the measurement.
In the reduced fit, all parameter shifts are below $1.2$ times
the corresponding symmetrized MINOS uncertainty of the full fit.
In particular, $T$, $C$, $\delta_0$, $\delta'_0$, $\delta_1$,
and $\phi$, which determine the CP-conserving neutral-$D$
amplitudes, shift by less than $0.4$ times the corresponding
full-fit uncertainties. The opposite-sign residuals in
$D^+\to\pi^0K^+$ and $D_s^+\to\pi^0K^+$ indicate competing
preferences for the common parameter $K'$. Allowing independent
breaking parameters for the two charged-meson sectors could reduce
these residuals, at the cost of an additional violation of the
corresponding $SU(3)_F$ relation. We retain a common $K'$ and
interpret the remaining discrepancy as a limitation of the adopted
$SU(3)_F$-breaking parametrization.

At the minimum of the full fit, the physical $\Delta U=1$
amplitudes for the four neutral-$D$ channels are (in GeV)
\begin{align}
\mathcal{A}(D^0 \to \pi^+ \pi^-)
&=
\bigl(-1.577 + 4.406\,i\bigr)\times 10^{-7},
\label{eq:app}\\
\mathcal{A}(D^0 \to K^+ K^-)
&=
\bigl(+5.898 - 5.949\,i\bigr)\times 10^{-7},
\label{eq:akk}\\
\mathcal{A}(D^0 \to K^0 \bar{K}^0)
&=
\bigl(+2.233 - 0.1732\,i\bigr)\times 10^{-7},
\label{eq:akk0}\\
\mathcal{A}(D^0 \to \pi^0 \pi^0)
&=
\bigl(+2.318 + 4.405\,i\bigr)\times 10^{-7}.
\label{eq:api0pi0}
\end{align}
where, for these SCS modes,
\begin{equation*}
\mathcal{A}(D \to f ) = \frac{G_F}{\sqrt{2}}\,\frac{1}{2}\left( V^{*}_{cs}V_{us} - V^{*}_{cd}V_{ud}\right) A(D \to f)\,.
\end{equation*}
In the absence of FSI, the corresponding reduced amplitudes simplify to
\begin{align*}
A(D^0\to K^+K^-)&=-A(D^0\to\pi^+\pi^-)=T,\\
A(D^0\to K^0\bar K^0)&=0,\\
A(D^0\to\pi^0\pi^0)&=C.
\end{align*}

\section{\texorpdfstring{The  $\Delta U = 0$ amplitudes and the CP violating asymmetries}
{The  Delta U = 0 amplitudes and the CP violating asymmetries}} 
\label{sec:DeltaU0}

The $\Delta U = 0$ amplitudes involve two additional reduced amplitudes, which we parametrize in terms of $\tilde{P}$ and $\Delta_4$.
They are associated with the singlet and octet components of
$
\bar{\mathbf{3}} \otimes \mathbf{3} = \mathbf{1} \oplus \mathbf{8},
$
while the reduced matrix elements associated with the $\mathbf{8}$ and $\mathbf{27}$ components of
$
\bar{\mathbf{3}} \otimes \mathbf{15}
$
are fixed by the branching-ratio fit. The final--state interactions are described by the same five parameters introduced for the 
$\Delta U = 1$ amplitudes and determined from the fit to the well-measured branching ratios.

We define the total $\Delta U = 0$ contribution of the penguin and
pseudopenguin operators to the physical decay amplitude for
$D^0\to f$ as
\begin{equation}
\mathcal{B}(D^0 \rightarrow f)
\equiv
\frac{G_F}{\sqrt{2}}
\left(
-\frac{V^\ast_{cb} V_{ub}}{2}
\right)
B(D^0 \rightarrow f),
\label{equation16}
\end{equation}
where $B(D^0\to f)$ denotes the corresponding hadronic amplitude. Their expressions are the following
\begin{align}
B(D^0\to\pi^+\pi^-) &= {\cal P} \left(\frac{1}{2} \left(e^{i\delta_0^\prime}
+ e^{i\delta_0}\right) + \left(e^{i\delta_0^\prime} - e^{i\delta_0}\right)  
\left(-\frac{1}{6} \cos(2\phi) - \frac{7}{4 \sqrt{10}} \sin(2\phi)
\right)\right) \nonumber\\
&+\left(T + C\right) \left(-\frac{3}{20}  \left(e^{i\delta_0^\prime} + 
e^{i\delta_0}\right) + \frac{3}{10} + \left(\frac{1}{60}  \cos(2\phi) + 
\frac{1}{2\sqrt{10}}  \sin(2\phi) \right) \left(e^{i\delta_0^\prime} - 
e^{i\delta_0} \right)\right)  \nonumber\\
&+\Delta_4  \left(e^{i\delta_0^\prime} - e^{i\delta_0}\right)  
\left(-\frac{1}{3} \cos(2\phi) - \frac{1}{4\sqrt{10}} \sin(2\phi)
\right),\nonumber\\
B(D^0\to\pi^0\pi^0) &= B(D^0\to\pi^+\pi^-) - (T + C),\nonumber\\
B(D^0\to K^+K^-) &= {\cal P}\left(\frac{1}{4}  \left(e^{i\delta_0^\prime} 
+ e^{i\delta_0}\right) + \left(e^{i\delta_0^\prime} - e^{i\delta_0}\right) 
\left(-\frac{5}{12}  \cos(2\phi) + \frac{1}{4\sqrt{10}} \sin(2\phi)\right) + 
\frac{1}{2} e^{i\delta_1}\right) \nonumber\\
& +\left(T + C\right)  \left(-\frac{1}{20}  \left(e^{i\delta_0^\prime} + 
e^{i\delta_0}\right) + \frac{3}{10} + \frac{7}{60}  \cos(2\phi) 
\left(e^{i\delta_0^\prime} - e^{i\delta_0}\right) - \frac{1}{5} e^{i\delta_1}
\right) \nonumber\\
& +\Delta_4  \left(\frac{1}{4}  \left(e^{i\delta_0^\prime} + e^{i\delta_0}
\right) + \left(e^{i\delta_0^\prime} - e^{i\delta_0}\right) 
\left(-\frac{1}{12}  \cos(2\phi) + \frac{3}{4\sqrt{10}}  \sin(2\phi)\right) - 
\frac{1}{2}  e^{i\delta_1}\right),\nonumber\\
B(D^0\to K^0\bar{K}^0) &= {\cal P}\left(\frac{1}{4} 
\left(e^{i\delta_0^\prime} + e^{i\delta_0}\right) + \left(e^{i\delta_0^\prime}
- e^{i\delta_0}\right)  \left(-\frac{5}{12} \cos(2\phi) + \frac{1}{4\sqrt{10}}
\sin(2\phi)\right) - \frac{1}{2} e^{i\delta_1}\right) \nonumber\\
&+ \left(T + C\right) \left(-\frac{1}{20} \left(e^{i\delta_0^\prime} + 
e^{i\delta_0}\right) - \frac{1}{10} + \frac{7}{60} \cos(2\phi) 
\left(e^{i\delta_0^\prime} - e^{i\delta_0}\right) + \frac{1}{5}  
e^{i\delta_1}\right) \nonumber\\
&+\Delta_4 \left(\frac{1}{4}  \left(e^{i\delta_0^\prime} + 
e^{i\delta_0}\right) + \left(e^{i\delta_0^\prime} - e^{i\delta_0}\right)  \left(-\frac{1}{12} \cos(2\phi) + \frac{3}{4\sqrt{10}} \sin(2\phi)\right) + \frac{1}{2}  e^{i\delta_1}\right).
\label{e:exprB}
\end{align}
where ${\cal P}=T+\widetilde P$ and
$\widetilde P\equiv P+\Delta_3$ is an effective reduced amplitude.
The contributions $P$ and $\Delta_3$ enter only through their sum
and cannot be disentangled within the present analysis.
The CP violating asymmetries are given by
\footnote{%
The decay amplitudes are written as
\begin{equation*}
\mathcal{M}(D\to f)=\frac{G_F}{\sqrt{2}}
\frac12\Big[\big(V_{cs}^{*}V_{us}-V_{cd}^{*}V_{ud}\big)\,A(D\to f)
-\big(V_{cb}^{*}V_{ub}\big)\,B(D\to f)\Big].
\end{equation*}
}%
\begin{equation}
A_{CP}(D\to f)\simeq
\Im\!\left[
  \,\frac{2\,V^{*}_{cb}V_{ub}}
         {V^{*}_{cs}V_{us} - V^{*}_{cd}V_{ud}}
\right]\,
\Im\left[\frac{B(D\to f)}{A(D\to f)}\right]\,,
\label{eq:ACP_CKM_yourconv}
\end{equation}
where
\begin{equation}
\Im\!\left[
  \frac{2\,V^{*}_{cb}V_{ub}}
         {V^{*}_{cs}V_{us} - V^{*}_{cd}V_{ud}}
\right]
= -6.446\times10^{-4} \,.
\label{equation18}
\end{equation}
The asymmetries calculated here are direct CP asymmetries.
For the $K_S^0K_S^0$ and $\pi^0\pi^0$ final states, we compare
our predictions with the available time-integrated measurements,
neglecting the indirect contribution in these comparisons.

Assuming $SU(3)_F$ symmetry for the non--rescattered amplitudes and adopting 
the pattern of final--state interactions extracted from the $\Delta U = 1$ 
sector, the $\Delta U = 0$ amplitudes can be decomposed into three $U$--spin 
singlets transforming as the $SU(3)$ representations $\mathbf{1}$, 
$\mathbf{8}$, and $\mathbf{27}$, with reduced matrix elements $A_{1}$, $A_{8}$ 
and $A_{27}$. 
The 27--plet contribution is fixed by the combination $T + C$ determined from 
branching ratios,
\begin{equation}
A_{27} = \sqrt{\frac{5}{6}}\,(T + C)\,.
\end{equation}
The remaining singlet and octet pieces can be expressed in terms of 
${\cal P}$, $\Delta_{4}$, and $T + C$ as 
\begin{align}
A_{1} &= \frac{\sqrt{2}}{3}\,\big[\,4{\cal P} + 2\Delta_{4} - (T + C)
\big]\,, \\
A_{8} &= \frac{2\sqrt{5}}{3}\,\big[\,{\cal P} - \Delta_{4} - \tfrac{2}{5}(T +
C)\big]\,.
\end{align}
For the non-rescattered amplitudes one has the $SU(3)_F$
relationships:
\begin{align} 
B_{nr}(D^0 \rightarrow \pi^+  \pi^-)  & = 
B_{nr}(D^0 \rightarrow K^+ K^-) 
\label{equation20}\\
B_{nr}(D^0 \rightarrow \pi^+  \pi^-) &  =  
B_{nr}(D^0 \rightarrow \pi^0  \pi^0) + T + C
\label{equation21} 
\end{align}
Using the numerical values obtained from the branching-ratio fit for all
parameters except for $\widetilde P$ and $\Delta_4$, and taking into account
final--state interactions, we obtain the following CKM-suppressed physical
amplitudes defined in Eq.~(\ref{equation16}).
The amplitudes $\mathcal{B}$ are given in GeV, with $\widetilde P$ and $\Delta_4$
understood as their numerical values in units of $\mathrm{GeV}^3$:
\begin{align}
\mathcal{B}(D^0 \!\to\! \pi^+ \pi^-)
&=
10^{-10}\left[
(1.706-0.9952\,i)
+(0.5013-2.791\,i)\,\Delta_{4}
+(5.171-3.628\,i)\,\widetilde{P}
\right],
\label{eq:pi+pi-}\\[5pt]
\mathcal{B}(D^0 \!\to\! K^+ K^-)
&=
10^{-10}\left[
(1.593+0.4322\,i)
-(0.2320+1.062\,i)\,\Delta_{4}
+(4.903+0.2171\,i)\,\widetilde{P}
\right],
\label{eq:K+K-}\\[5pt]
\mathcal{B}(D^0 \!\to\! K^0 \bar{K}^0)
&=
10^{-10}\left[
(0.2786-1.611\,i)
+(4.318+3.472\,i)\,\Delta_{4}
+(0.3527-4.317\,i)\,\widetilde{P}
\right],
\label{eq:K0K0bar}\\[5pt]
\mathcal{B}(D^0 \!\to\! \pi^0 \pi^0)
&=
10^{-10}\left[
(2.287-2.251\,i)
+(0.5013-2.791\,i)\,\Delta_{4}
+(5.171-3.628\,i)\,\widetilde{P}
\right].
\label{eq:pi0pi0}
\end{align}
which make it possible to express the CP–violating asymmetries in terms of the
parameters determined from the fit to the branching ratios discussed in the 
previous section, and of the two remaining free parameters, $\tilde{P}$ and 
$\Delta_4$.
In general, we can write  
\begin{equation}
\label{eq:lin-expansion}
A_{CP}(f)\;\simeq\; a_f \;+\; b_f\,\tilde P \;+\; c_f\,\Delta_4 \,,
\end{equation}
where \((a_f, b_f, c_f)\) are channel--dependent real coefficients determined 
by the hadronic parameters  
(\(T, C, \kappa, \kappa', K, K'\), strong phases, etc.) once these are fixed by 
the fit to the branching ratios.  So we have
%
%
\begin{align}
A_{CP}(D^0 \to \pi^+\pi^-)
&\simeq
10^{-3}\left(
-0.7390 - 0.9077\,\Delta_4 - 2.409\,\widetilde{P}
\right),
\label{e:CPPDelta1lin}
\\[2mm]
A_{CP}(D^0 \to K^+K^-)
&\simeq
10^{-3}\left(
+0.2909 + 0.03299\,\Delta_4 + 0.9793\,\widetilde{P}
\right),
\label{e:CPPDelta2lin}
\\[2mm]
A_{CP}(D^0 \to K_S^0K_S^0)
&\simeq
10^{-3}\left(
-0.2439 + 4.257\,\Delta_4 - 0.9516\,\widetilde{P}
\right),
\label{e:CPPDelta3lin}
\\[2mm]
A_{CP}(D^0 \to \pi^0\pi^0)
&\simeq
10^{-3}\left(
-0.7771 - 1.006\,\Delta_4 - 1.225\,\widetilde{P}
\right),
\label{e:CPPDelta4lin}
\end{align}
which imply the constraints among the four CP-violating asymmetries that we 
shall discuss in the following.  

Therefore, our model implies quantitative relations among the different
CP asymmetries. In particular, the asymmetries of the neutral final states
can be expressed in terms of the two charged-mode asymmetries measured in
\(D^0\to K^+K^-\) and \(D^0\to\pi^+\pi^-\) decays. Defining
\begin{equation}
\Sigma A_{CP}
\equiv
A_{CP}(K^+K^-)+A_{CP}(\pi^+\pi^-),
\qquad
\Delta A_{CP}
\equiv
A_{CP}(K^+K^-)-A_{CP}(\pi^+\pi^-),
\end{equation}
we obtain
\begin{align}
A_{CP}(D^0 \to K_S^0K_S^0)
&=
-8.275\times10^{-5}
-9.462\,\Sigma A_{CP}
-4.274\,\Delta A_{CP},
\label{eq:ACPKsKs-sigma-delta}\\[3mm]
A_{CP}(D^0 \to \pi^0\pi^0)
&=
-3.859\times10^{-4}
+1.395\,\Sigma A_{CP}
+0.2270\,\Delta A_{CP}.
\label{eq:ACPpi0pi0-sigma-delta}
\end{align}

Using the LHCb measurements reported in Ref.~\cite{LHCb:2022},
see Eqs.~(\ref{e:LHCbpipi}) and~(\ref{e:LHCbKK}), and taking into account
their correlation coefficient $\rho=0.88$, the relations in
Eqs.~(\ref{eq:ACPKsKs-sigma-delta}) and~(\ref{eq:ACPpi0pi0-sigma-delta})
give
\begin{align}
A_{CP}(D^0 \to K_S^0 K_S^0)
& =  (-22.7 \pm 10.7)\times 10^{-3}\,,\\
A_{CP}(D^0 \to \pi^0 \pi^0)
& =  (+3.57 \pm 1.59)\times 10^{-3}\,.
\end{align}
The two predictions are almost fully anticorrelated,
$\rho\left[A_{CP}(K_S^0K_S^0),A_{CP}(\pi^0\pi^0)\right]\simeq-0.997$,
since both are driven by the same pair of LHCb inputs.
The prediction for the \(\pi^0\pi^0\) mode can already be compared with
the Belle~II measurement quoted in Eq.~(\ref{eq:BelleII-pi0pi0}).
The predicted and measured central values are very close and fully
compatible within the present uncertainties.  However, the experimental
uncertainty is still too large to provide a significant constraint on
\(\tilde{P}\) and \(\Delta_4\).

Moreover, we determine the parameters $\tilde{P}$ and $\Delta_4$ 
by solving Eqs.~(\ref{e:CPPDelta1lin}) and~(\ref{e:CPPDelta2lin}) with the
measurements reported in Ref.~\cite{LHCb:2022}. Including the correlation 
coefficient \(\rho=0.88\), 
we find
\begin{align}
\tilde{P} &= +0.662 \pm 0.661, \\
\Delta_4  &= -5.13 \pm 2.37.
\end{align}
with correlation coefficient, $
\rho\simeq -0.99$.
This strong anti-correlation shows that the two charged-mode asymmetries do not
determine \(\tilde{P}\) and \(\Delta_4\) independently: a variation of
one parameter can be partially compensated by a correlated variation of
the other one. Therefore, the separate errors on \(\tilde{P}\) and
\(\Delta_4\) should not be interpreted as independent uncertainties.

In particular, \(\tilde{P}\) is compatible with zero at approximately
one standard deviation, whereas the extracted value of \(\Delta_4\)
shows a moderate preference for a nonzero value, at the level of about
\(2.2\sigma\). The charged-mode asymmetries therefore favour a
\(\Delta_4\)-like contribution, although the present significance is not
sufficient to establish it conclusively. Moreover, the strong
anti-correlation between \(\tilde{P}\) and \(\Delta_4\) prevents a clean
separation of the two contributions. Additional independent information,
for instance from a neutral-final-state CP asymmetry with a different
dependence on the two parameters, would help break this degeneracy and
constrain the two components separately. Complementary theoretical input
may also help identify the phenomenologically preferred region of
parameter space.

\section{An OZI-motivated benchmark for the CP asymmetries}
\label{sec:OZI-benchmark}

The two-parameter analysis discussed above shows that the CP asymmetries
in the charged final states constrain the parameters \(\tilde{P}\) and
\(\Delta_4\) with a strong anti-correlation.  Consequently, the modes
with neutral final states play a crucial role in discriminating among
different dynamical scenarios.  In this section, we consider a
particularly constrained benchmark motivated by the Zweig, or OZI,
selection rule~\cite{Okubo:1963fa,Zweig:1964jf,Iizuka:1966fk}.

The Zweig/OZI rule suppresses disconnected, non--rescattered matrix
elements.  In our parametrization, this implies, in particular,
\begin{equation}
B_{\rm nr}(D^0\to K^0\bar K^0)=0\,.
\end{equation}
Using the expression for \(B_{\rm nr}(D^0\to K^0\bar K^0)\) given in
Eq.~(\ref{e:exprB}), this condition directly yields
\begin{equation}
\Delta_4=0\,.
\end{equation}
The CP asymmetries then depend on the single remaining parameter
\(\tilde{P}\).  To retain the modes \(D^0\to K_S^0 K_S^0\) and
\(D^0\to\pi^0\pi^0\) as genuine predictions of this benchmark, we
determine \(\tilde{P}\) using only \(\Delta A_{CP}\).
For \(\Delta_4=0\), the expressions obtained above give
\begin{equation}
\Delta A_{CP}
=
\left(1.030+3.388\,\tilde{P}\right)\times10^{-3}.
\end{equation}
Using the experimental input
\begin{equation}
\Delta A_{CP}^{\rm exp}
=
(-1.55\pm0.29)\times10^{-3},
\end{equation}
we obtain
\begin{equation}
\tilde{P}=-0.761\pm0.086\,.
\end{equation}
The resulting predictions for the modes with neutral final states are
\begin{align}
A_{CP}(D^0\to K_S^0 K_S^0)
&=
(0.481\pm0.082)\times10^{-3},
\\
A_{CP}(D^0\to\pi^0\pi^0)
&=
(0.155\pm0.105)\times10^{-3}.
\end{align}
For completeness, the corresponding values for the charged-final-state
modes are
\begin{align}
A_{CP}(D^0\to\pi^+\pi^-)
&=
(1.10\pm0.21)\times10^{-3},
\\
A_{CP}(D^0\to K^+K^-)
&=
(-0.455\pm0.084)\times10^{-3}.
\end{align}
Since \(\Delta A_{CP}\) is used to determine \(\tilde{P}\), the predicted
asymmetries in the \(\pi^+\pi^-\) and \(K^+K^-\) channels necessarily
reproduce their measured difference.  Their individual measurements
nevertheless provide a further consistency test of this benchmark.

The quoted uncertainties arise solely from the experimental uncertainty
on \(\Delta A_{CP}\), propagated through the determination of
\(\tilde{P}\).  They do not yet include the uncertainties and
correlations of the coefficients obtained from the branching-fraction
fit. 

\begin{figure}[t]
\centering
\includegraphics[width=0.80\textwidth]{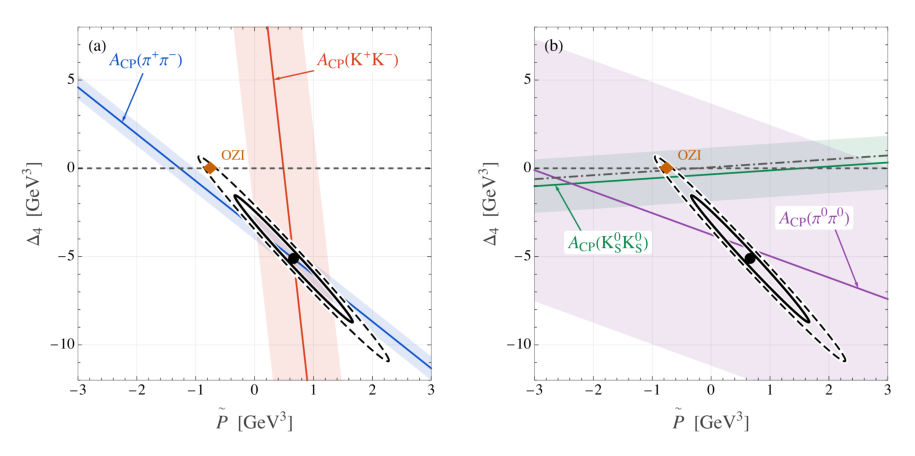}
\caption{Constraints in the $(\widetilde P,\Delta_4)$ plane from the charged-mode
direct CP asymmetries in panel~(a) and from the available time-integrated
neutral-mode asymmetries in panel~(b), the latter being interpreted as
direct asymmetries after neglecting the indirect contribution at the
present experimental precision. The hadronic coefficients are fixed at
their best-fit values from the branching-fraction analysis. The coloured
lines and bands denote the central values and individual $1\sigma$
intervals. The black solid and dashed contours are the conditional
$68.3\%$ and $95\%$ confidence regions obtained from the correlated
charged-mode measurements. The black dot marks the best-fit point,
whereas the orange diamond denotes the fit constrained by
$\Delta_4=0$. The dot-dashed line in panel~(b) corresponds to
$a_{CP}^{\rm dir}(D^0\to K_S^0K_S^0)=0$.
}
\label{fig:ptilde-delta4-plane}
\end{figure}

Figure~\ref{fig:ptilde-delta4-plane} displays the constraints in the
$(\widetilde P,\Delta_4)$ plane, with the hadronic coefficients fixed
at their best-fit values from the branching-ratio analysis.
For each channel, the coloured line reproduces the experimental
central value, and the shaded band represents its individual
$\pm1\sigma$ uncertainty. The arrows identify the corresponding
asymmetries.

The black contours are obtained from a fit to
$A_{CP}(D^0\to\pi^+\pi^-)$ and $A_{CP}(D^0\to K^+K^-)$, including
their experimental correlation $\rho=0.88$.
The solid and dashed contours correspond to
$\Delta\chi^2=2.30$ and $5.99$, respectively, defining the
$68.3\%$ and $95\%$ confidence regions for two parameters.
The black dot marks the best fit. The neutral-final-state
asymmetries are shown for comparison and are not included in this fit.

The grey dashed line denotes the OZI limit, $\Delta_4=0$,
while the grey dot-dashed line corresponds to
$A_{CP}(D^0\to K_S^0K_S^0)=0$; these are distinct conditions.
The orange diamond marks the minimum of the same correlated
two-input $\chi^2$ with $\Delta_4$ fixed to zero.

\section{Conclusions}
\label{sec:conclusions}

We have updated the $SU(3)_F$ analysis of $D\to PP$ decay amplitudes,
including final--state interactions and the most recent experimental
information on the branching fractions.  The present fit provides a
substantially improved description of the updated PDG(2026) data
~\cite{PDG2026} with respect to the previous analysis.  The remaining
discrepancies are concentrated in a small number of modes involving
charged charmed mesons, while all the neutral-$D$ SCS branching fractions
relevant for the analysis of CP violation are reproduced within
approximately $2\sigma$.

Once the CP-conserving hadronic parameters are fixed by the
branching-ratio fit, the direct CP asymmetries depend on the two additional
parameters $\tilde{P}$ and $\Delta_4$.  Constraining them with the
measured charged-mode asymmetries leads, in the general two-parameter
analysis, to
\begin{align}
A_{CP}(D^0 \to K_S^0 K_S^0)
& =  (-22.7 \pm 10.7)\times 10^{-3}\,,\\
A_{CP}(D^0 \to \pi^0 \pi^0)
& =  (+3.57 \pm 1.59)\times 10^{-3}\,.
\end{align}
The latter is fully compatible with the present Belle~II measurement,
while the percent-level prediction for the $K_S^0 K_S^0$ asymmetry provides
a particularly sensitive test of this scenario.

A particularly interesting limiting case is obtained by imposing the
Zweig/OZI condition on the non--rescattered amplitudes, which in our
parametrization corresponds to $\Delta_4=0$.  The CP-violating sector
then depends on the single parameter $\tilde{P}$, and using the
measured $\Delta A_{CP}$ we obtain
\begin{align}
A_{CP}(D^0\to K_S^0 K_S^0)
&=
(0.481\pm0.082)\times10^{-3},
\\
A_{CP}(D^0\to\pi^0\pi^0)
&=
(0.155\pm0.105)\times10^{-3}.
\end{align}
The same limit predicts opposite signs for the
$D^0\to\pi^+\pi^-$ and $D^0\to K^+K^-$ asymmetries.
As shown in Fig.~\ref{fig:ptilde-delta4-plane}, the OZI condition
$\Delta_4=0$ is distinct from the phenomenological condition
$A_{CP}(D^0\to K^0_S K^0_S)=0$.

The present experimental uncertainty on
$A_{CP}(D^0\to K^0_S K^0_S)$ is still too large to discriminate decisively
between the different possibilities.   More precise measurements of
$A_{CP}(D^0\to K^0_S K^0_S)$ and $A_{CP}(D^0\to\pi^0\pi^0)$ would therefore
be especially valuable, since their different dependence on
$\tilde{P}$ and $\Delta_4$ can break the strong anti-correlation left
by the charged-mode data.  A more precise determination of the sum of
the charged-mode asymmetries would provide a complementary test of the
opposite-sign pattern predicted in the OZI limit.

\section*{Acknowledgments}
The authors acknowledge the use of ChatGPT (OpenAI) and Claude (Anthropic) for 
manuscript editing and coding assistance.

\appendix

\section{\texorpdfstring{The $\Delta U = 1$ Amplitudes}{The \Delta U = 1 Amplitudes}}
\label{sec:DeltaU=1A}

Here we report the full expressions for the $\Delta U = 1$ amplitudes \cite{BPS}.
{\allowdisplaybreaks
\begin{center}
{\bf SCS modes} (to be multiplied by $\frac{1}{2}(V^{*}_{cs}V_{us} - V^{*}_{cd}V_{ud})$)
\end{center}
\begin{align}
 A(D^0\to \eta_8\eta_8) &= \left(T - \frac{2}{3} C\right) \left[\frac{3}{10} \left(e^{i \delta_0} + e^{i \delta_0^\prime}\right) + \left(\frac{3}{10}  \cos(2 \phi) + \frac{3}{4\sqrt{10}} \sin(2 \phi)\right)  \left(e^{i \delta_0^\prime} - e^{i \delta_0}\right)\right]\nonumber\\& -  \frac{3}{5}\left(T + C\right)\nonumber\\
A(D^0\to \pi^0\eta_8) &= \frac{\sqrt{3}}{5} \left[\left(T - \frac{2}{3} C\right)  e^{i\delta_1} - \left(T + C\right)\right]\nonumber\\
A(D^+\to K^+\bar{K}^0) &= \frac{1}{5}  (2  T - 3 C + \Delta)  e^{i \delta_1} + \frac{3}{5} (T + C)\nonumber\\
A(D^+\to \pi^+\pi^0) &= \frac{1}{\sqrt{2}} (T + C)\nonumber\\
A(D^+\to \pi^+\eta_8) &= \frac{\sqrt{2}}{5\sqrt{3}}  (2  T - 3 C + \Delta)  e^{i \delta_1} - \frac{3\sqrt{3}}{5\sqrt{2}} (T + C)\nonumber\\
A(D_s^+\to \pi^+K^0) &= -\frac{1}{5}  (2  T - 3 C + \Delta - K^\prime)  e^{i \delta^\prime_{\frac{1}{2}}} - \frac{3}{5} (T + C)\nonumber\\
A(D_s^+\to \pi^0K^+) &= -\frac{1}{5\sqrt{2}}  (2  T - 3 C + \Delta - K^\prime)  e^{i \delta^\prime_{\frac{1}{2}}} + \frac{2}{5\sqrt{2}} (T + C)\nonumber\\
A(D_s^+\to K^+\eta_8) &= \frac{1}{5\sqrt{6}}  (2  T - 3 C + \Delta - K^\prime)  e^{i \delta^\prime_{\frac{1}{2}}} - \frac{2\sqrt{6}}{5} (T + C)
\label{eq:SCS_P}
\end{align}
\begin{center}
{\bf CA modes} (to be multiplied by $V^{*}_{cs}V_{ud}$)
\end{center}
\begin{align}
A(D^0\to \pi^+K^-) &= \frac{1}{5}  (3  T - 2 C - K)  e^{i \delta_{\frac{1}{2}}} + \frac{2}{5} (T + C + \kappa)\nonumber\\
A(D^0\to \pi^0\bar{K}^0) &= -\frac{1}{5\sqrt{2}}  (3  T - 2 C - K)  e^{i \delta_{\frac{1}{2}}} + \frac{3}{5\sqrt{2}} (T + C + \kappa)\nonumber\\
A(D^0\to \bar{K}^0 \eta_8) &= -\frac{1}{5\sqrt{6}}  (3  T - 2 C - K)  e^{i \delta_{\frac{1}{2}}} + \frac{3}{5\sqrt{6}} (T + C + \kappa)\nonumber\\
A(D^+ \to \pi^+\bar{K}^0)  &= (T + C+ \kappa)\nonumber\\
A(D_s^+\to K^+\bar{K}^0) &= -\frac{1}{5}  (2  T - 3 C + \Delta)  e^{i \delta^\prime_{1}} + \frac{2}{5} (T + C + \kappa)\nonumber\\
A(D_s^+\to \pi^+\eta_8) &= -\frac{\sqrt{2}}{5\sqrt{3}}  (2  T - 3 C + \Delta)  e^{i \delta^\prime_{1}} - \frac{\sqrt{6}}{5} (T + C + \kappa)
\label{eq:CA_P}
\end{align}
\begin{center}
{\bf DCS modes} (to be multiplied by $-V^{*}_{cd}V_{us}$)
\end{center}
\begin{align}
A(D^0\to \pi^0 K^0) &= \frac{1}{5\sqrt{2}}  (3  T - 2 C + K)  e^{i \delta_{\frac{1}{2}}} - \frac{3}{5\sqrt{2}} (T + C + \kappa^\prime)\nonumber\\
A(D^0\to K^0 \eta_8) &= \frac{1}{5\sqrt{6}}  (3  T - 2 C + K)  e^{i \delta_{\frac{1}{2}}} - \frac{3}{5\sqrt{6}} (T + C + \kappa^\prime)\nonumber\\
A(D^+ \to \pi^+K^0) &= \frac{1}{5}  (2  T - 3 C + \Delta - K^\prime)  e^{i \delta_{\frac{1}{2}}} - \frac{2}{5} (T + C + \kappa^\prime)\nonumber\\
A(D^0\to \pi^- K^+) &= -\frac{1}{5}  (3  T - 2 C + K)  e^{i \delta_{\frac{1}{2}}} - \frac{2}{5} (T + C + \kappa^\prime)\nonumber\\
A(D^+ \to \pi^0K^+) &= \frac{1}{5\sqrt{2}}  (2  T - 3 C + \Delta - K^\prime)  e^{i \delta_{\frac{1}{2}}} + \frac{3}{5\sqrt{2}} (T + C + \kappa^\prime)\nonumber\\
 A(D^+ \to K^+\eta_8) &= -\frac{1}{5\sqrt{6}}  (2  T - 3 C + \Delta - K^\prime)  e^{i \delta_{\frac{1}{2}}} - \frac{3}{5\sqrt{6}} (T + C + \kappa^\prime)\nonumber\\
A(D_s^+ \to K^+K^0)  &= -(T + C + \kappa^\prime)
\label{eq:DCS_P}
\end{align}
}

\end{document}